\documentclass[aps,nofootinbib,preprintnumbers,reprint]{revtex4-1}
\usepackage{geometry}     
\usepackage{graphicx}
\usepackage{amsmath,amssymb}  
\usepackage{bm}  
\usepackage{slashed}
\usepackage{color}
\usepackage{subfig}
\newcommand\beq{\begin{equation}}
\newcommand\eeq{\end{equation}}

\newcommand\nn{\nonumber}

\newcommand\sfrac[2]{{\textstyle{\frac{#1}{#2}}}}

\def\beq{\begin{eqnarray}}
\def\eeq{\end{eqnarray}}

\def\ba{\begin{eqnarray}}
\def\ea{\end{eqnarray}}
\def\beq{\begin{eqnarray}}
\def\eeq{\end{eqnarray}}

\def\L*{{\cal L}_*}
\def\L{\mathcal{L}}
\def\({\left(}
\def\){\right)}

\def\nn{\nonumber}

\def\<{\langle}
\def\>{\rangle}

\newcommand{\be}{\begin{equation}}
\newcommand{\ee}{\end{equation}}

\newcommand{\bea}{\begin{eqnarray}}
\newcommand{\eea}{\end{eqnarray}}
\newcommand{\beas}{\begin{eqnarray*}}
\newcommand{\eeas}{\end{eqnarray*}}

\def\({\left(}
\def\){\right)}

\def\lsim{\mathrel{\rlap{\lower3pt\hbox{\hskip0pt$\sim$}}
     \raise1pt\hbox{$<$}}}         
\def\gsim{\mathrel{\rlap{\lower4pt\hbox{\hskip1pt$\sim$}}
     \raise1pt\hbox{$>$}}}         

\def\lsim{\mathrel{\rlap{\lower3pt\hbox{\hskip0pt$\sim$}}
     \raise1pt\hbox{$<$}}}         
\def\gsim{\mathrel{\rlap{\lower4pt\hbox{\hskip1pt$\sim$}}
     \raise1pt\hbox{$>$}}}         

\begin{document}
\title{CP Violation and Flavor SU(3) Breaking in D-meson Decays}


\author{David Pirtskhalava}
\email[Electronic address: ]{pirtskhalava@physics.ucsd.edu}
\affiliation{Department of Physics, University of California at San Diego, La Jolla, CA 92093}

\author{Patipan Uttayarat}
\email[Electronic address: ]{puttayarat@physics.ucsd.edu}
\affiliation{Department of Physics, University of California at San Diego, La Jolla, CA 92093}

\preprint{UCSD/PTH 11-21}
\begin{abstract}
We carry out a systematic flavor $SU(3)$ analysis of D-meson decays including the leading order symmetry breaking effects. We find that $SU(3)$ breaking can easily account for the recent LHCb measurement of the difference in CP asymmetries in the decays of $D^0$ into $K^+K^-$ and $\pi^+\pi^-$ mesons, once an enhancement mechanism, similar to the $\Delta=1/2$ rule in neutral kaon decays is assumed. As a byproduct of the analysis, one can make predictions regarding the individual asymmetries in $K^+K^-$, $\pi^+\pi^-$, as well as the $D^0 \to \pi^0\pi^0$ decay channels. Moreover, we find that the asymmetry in the decay $D^+\to \pi^+\pi^0$ vanishes in the leading approximation.
\end{abstract}

\maketitle

\section{Introduction and Summary}
\label{sec:intro}
It is commonly believed that the amount of CP violation (CPV)  in $D$ decays is small within the Standard Model (SM) and any appreciable CPV effects would be an indication of new physics. Nevertheless, what is meant by `small' is uncertain because we lack tools to reliably calculate matrix elements in QCD. 

Recently the LHCb collaboration released a 3.5$\sigma$ evidence for the difference between the time-integrated CP asymmetries in the two $D^0$ meson decay modes, $D^0\to K^+K^-$ and $D^0\to \pi^+\pi^-$~\cite{Aaij:2011in},
\begin{multline}
\label{lhcbresult}
	A_{CP}(D^0 \to K^+K^-)- A_{CP}(D^0 \to \pi^+\pi^-)\\=-(0.82\pm 0.21\pm0.11)\%,
\end{multline}
where the asymmetry in the decay into a final state $f$ is defined as follows
\beq
A_{CP}(D^0 \to f)=\frac{\Gamma (D^0\to f) - \Gamma (\bar{D^0}\to \bar f) }{\Gamma (D^0\to f) + \Gamma (\bar{D^0}\to \bar f)}.\nn
\eeq 

It was suggested in \cite{Brod:2011re}, that enhancement of certain formally $c$-quark-mass suppressed penguin diagrams can contribute significantly to the measured asymmetry, while an extensive effective field theoretic analysis of operators, that can possibly give rise to $\Delta A_{CP}$ has been performed in \cite{Isidori:2011qw}.

In this note we carry out a systematic flavor $SU(3)$  analysis of $D$-meson decays including the leading order symmetry breaking effects due to the nonzero strange quark mass $m_s$. 

The  flavor $SU(3)$ symmetry of quarks, although not an exact symmetry of the low-energy SM, has nevertheless proven to provide a powerful tool for extracting information about meson decays, more or less independent of the details of the strong interactions. An $SU(3)$ analysis of $D$ decay amplitudes, neglecting CP violation and the difference between the $s$ and $(u,d)$ quark masses, has first been carried out in Ref.~\cite{Quigg:1979ic}. In \cite{Golden:1989qx} the approach was extended to incorporate CPV effects, including also short distance QCD corrections. In the $SU(3)$ limit, the $D^0$ decay amplitudes into $ K^+K^-$ and $\pi^+\pi^-$ final states are given by
\beq
\begin{split}
 \mathcal{A}(D^0 \to K^+K^-)&=\tilde{a}\Sigma+\tilde{b}\Delta,\\
 \mathcal{A}(D^0 \to \pi ^+\pi ^-)&=-\tilde{a}\Sigma+\tilde{b}\Delta,
\end{split}
\label{su3limit}
\eeq
where $\Sigma$ and $\Delta$ denote certain combinations of the elements of the Cabibbo-Kobayashi-Maskawa (CKM) matrix, satisfying $|\Delta|\ll |\Sigma|$, while $\tilde{a}$ and $\tilde{b}$ represent particular strong interaction matrix elements described below. The problem with the above expression is that reconciling it with the observed ratio of the partial rates $\Gamma(D^0 \to K^+K^-)/\Gamma(D^0 \to \pi^+\pi^-)\simeq 2.8$ requires a large ($\sim$ three orders of magnitude) enhancement of the matrix element $\tilde{b}$ with respect to $\tilde{a}$. While significant enhancement can be motivated by empirical experience\footnote{Indeed, we will see below that $\tilde{b}$ contains a contribution from the matrix element of the $SU(3)$ triplet part of the weak Hamiltonian, while $\tilde{a}$ receives contributions only from higher representations. The enhancement of lower representation matrix elements similar to the $\Delta I=1/2$ rule in kaon physics is a well-known, though poorly understood experience.}, explaining this ratio in the $SU(3)$ limit would require an unacceptably large amount of CP violation in $D^0$ decays \cite{Golden:1989qx}. Already in \cite{Golden:1989qx} it was noted however that the most probable resolution of this puzzle is to assume appreciable $SU(3)$ breaking effects in the processes at hand.   

Below we present a detailed analysis of $SU(3)$ breaking in $D$ decays (see \cite{Savage:1991wu,Hinchliffe:1995hz,Grinstein:1996us} for related work). 
Under the fairly general assumptions that \textbf{a)}\textit{only leading symmetry breaking effects, e.g., first order in $m_s$, need be retained}, and \textbf{b)}\textit{lower $SU(3)$ representations of the weak Hamiltonian lead to somewhat enhanced hadronic matrix elements, much as in the $\Delta I=1/2$ rule in neutral kaon decays}, we find that the observed asymmetry can easily be reconciled with the measured values for the partial rates without an unacceptably large enhancement of matrix elements. As a byproduct, one can make a number of predictions about the individual asymmetries in the decays of $D^0$ into $K^+K^-$, $\pi^+\pi^-$, as well as $\pi^0\pi^0$ final states. 

In this  work we have ignored the effect of $\eta-\eta^\prime$ mixing since the amplitudes we are interested in do not involve the $\eta$ meson. A more complete analysis involving the $\eta-\eta'$ mixing will be presented elsewhere.

The letter is organized as follows. In Sec.~\ref{sec:su3} we briefly describe the general procedure of applying the $SU(3)$ analysis to relating amplitudes of different $D$ meson decay modes to each other; both the $SU(3)$ limit and symmetry-breaking expressions are derived.
Sec.~\ref{sec:decaywidth}. deals with the analysis of a particular subclass of $D$ decay channels that is of interest in light of the recent LHCb results; we show that all observed rates and CP asymmetries can easily be accommodated within the broken $SU(3)$ framework. In Sec.~\ref{sec:cpv}. we discuss the predictions of the framework, while all technical details including the complete table of $D$ decay amplitudes into two pseudo-scalar mesons are collected in the appendix.

\section{SU(3) Analysis}
\label{sec:su3}
We start with reviewing the group structure of hadronic weak currents. Following the notation of Quigg \cite{Quigg:1979ic}, the $\Delta C = -1$ Hamiltonian relevant for the analysis of $D$ meson decays is given by the following expression,
\begin{multline}
\mathcal{H}=T^{31}_2 V_{11}V_{22}^*+T^{21}_3V^{\phantom{\ast}}_{12}V_{21}^*\\
+(T^{31}_3-T^{21}_2)\Sigma+(T^{13}_3-T^{12}_2)\Delta.
\label{h}
\end{multline}
Here the tensor $T^{ij}_k \equiv \bar 3\times8=\bar 3+6+\overline{15}$ is defined (suppressing the $V-A$ structure for simplicity) in terms of the quark $SU(3)$ triplet $\psi$ as follows,
\beq
T^{ij}_k = (\bar{\psi}^i\psi_k)(\bar{\psi}^jc)-\frac{1}{3}\delta^{i}_k (\bar{\psi}^l\psi_l)(\bar{\psi}^jc),
\eeq
while the elements of the CKM matrix $V_{ij}$ and the quantities $\Sigma$ and $\Delta$ are given in the standard notation by the following expressions,
\beq
\begin{split}
\Sigma&=\frac{1}{2•}(V^{\phantom{\ast}}_{12}V_{22}^*- V^{\phantom{\ast}}_{11}V_{21}^*),\\
\Delta&=\frac{1}{2•}(V^{\phantom{\ast}}_{12}V_{22}^*+ V^{\phantom{\ast}}_{11}V_{21}^*).
\end{split}
\eeq
We use the standard CKM parameterization \cite{Nakamura:2010zzi} with $ \theta_{12}=0.23,~ \theta_{13}=0.003, ~\theta_{23}=0.04$, and $\delta=1.2$. Numerically, $|\Sigma|/|\Delta|\sim 3000$. 

Below we consider the decays of a $D$-meson into two pseudo-scalars. Bose statistics only allows for  symmetric final states, $f=(8\times8)_s= 1+8+27$,  and the weak Hamiltonian \eqref{h} can be written in terms of the different $SU(3)$ representations as follows,
\begin{equation}
\begin{split}
\mathcal{H}&=\(\frac{1}{2•}[\overline{15}]^{31}_2+\frac{1}{4•}[6]_{22}\)V^{\phantom{\ast}}_{11}V_{22}^*\\
&\quad +\frac{1}{•2}\([\overline{15}]^{31}_3-[\overline{15}]^{21}_2+[6]_{23}\)\Sigma \\
&\quad +\(-\frac{1}{2•}[\overline{15}]^{11}_1 +\frac{3}{4•}[\bar 3]^1\)\Delta\\
&\quad +\(\frac{1}{2•}[\overline{15}]^{21}_3-\frac{1}{4•}[6]_{33}\)V^{\phantom{\ast}}_{12}V_{21}^*,
\label{rep1}
\end{split}
\end{equation}
where the corresponding representations are defined in terms of the tensor $T$ in the following way, 
\beq
\begin{split}
[\bar 3]^{i\phantom{k}} &=T^{ij}_j,\\
[6]_{kl} &=\varepsilon_{kij}T^{ij}_l+\varepsilon_{lij}T^{ij}_k,\\
[\overline{15}]^{ij}_k&=T^{ij}_k+T^{ji}_k-\frac{1}{4•}\delta^i_kT^{jl}_l-\frac{1}{4•}\delta^j_kT^{il}_l.
\end{split}
\eeq
Here upper/lower indices correspond the fundamental/antifundamental representation of the flavor $SU(3)$ group.

Consider a matrix element $\langle f| O |i\rangle$ between an initial $D$ meson state $i_r$ and a final state $f^{ij...}_{mn...}$ of an operator $O^{xy\ldots}_{uv\ldots}$ contributing to the Hamiltonian and belonging to some definite representation of the symmetry group. Invariance under $SU(3)$ constrains the matrix elements to be of the form,
\beq
\langle f^{ij...}_{mn...}| O^{xy...}_{uv...} |i_r\rangle = M \mathcal{T}^{ij...xy...}_{mn...uv...r},
\label{rep}
\eeq 
where $\mathcal{T}$ represents a tensor made out of the invariant tensors $\delta$'s and $\varepsilon$'s \footnote{In general there can be more than one invariant tensor with its own reduced amplitude on the right hand side of \eqref{rep}.}, while $M$ denotes a reduced matrix element encoding all the strong dynamics of the system. Calculating the invariant tensor $\mathcal{T}$ for each group representation contributing to the Hamiltonian, one can use Eq.~\eqref{rep} to relate different matrix elements to each other.

Neglecting isospin breaking, one can parametrize the breaking of flavor $SU(3)$ symmetry in strong interactions by 
\beq
\Delta\mathcal{L}_{QCD} = -m_s\bar\psi\lambda^8\psi,
\eeq
with $\lambda^8$ being one of the two diagonal Gell-Mann matrices\footnote{This parameterization of $SU(3)$ breaking was used for obtaining the Gell-Mann-Okubo mass formulae \cite{GellMann:1961ky,Okubo:1961jc}.}. Incorporating the symmetry-breaking to first order in the strange quark mass, the $SU(3)$ structure of the weak Hamiltonian becomes,
\begin{equation}
\begin{split}
\mathcal{H} &=\(\bar 3+6+\overline{15}\)\times \(1+\epsilon ~ 8+\mathcal{O}(\epsilon^2)\)\\
&\supset \bar{3}+6+\overline{15}+\epsilon\left( \bar{3}_i+6_i+\overline{15}^{\phantom{b}}_1+\overline{15}^{\phantom{b}}_2\right. \\
&\qquad \left. +\overline{15}^1_3+\overline{15}^2_3+\overline {24}_3+\overline{42}_3 +\dots\right),
\end{split}
\end{equation}
where the subscript $i =$ (1,2,3) indicates which of the ($\bar{3}$, $6$, $\overline{15}$) representations in \eqref{rep1} the $SU(3)$ breaking operators are obtained from, while $\epsilon$ represents a formal parameter counting the order of $SU(3)$ breaking.
Note that $\overline{15}\times 8$ includes yet another dimension-fifteen representation which we ignore here since upon closer inspection its matrix elements between the desired states vanish by group theory.

The complete list of invariant amplitudes including the leading $SU(3)$ breaking operators is, 
\begin{equation}
\label{matrixelements}
\begin{aligned}
	\langle 1| 3_{(i)}| 3\rangle &= G_{(i)},\\
	\langle 8| 3_{(i)}| 3\rangle &= F_{(i)},\\
	\langle 8| 6_{(i)}| 3\rangle &= S_{(i)},\\
	\langle 8| \overline{15}^{(\alpha)}_{(i)}| 3\rangle &= E^{(\alpha)}_{(i)},
\end{aligned}\quad
\begin{aligned}
	\langle 27| \overline{15}^{(\alpha)}_{(i)}| 3\rangle &= T^{(\alpha)}_{(i)},\\
	\langle 27| \overline{24}_{(i)}| 3\rangle &= H_{(i)},\\
	\langle 27| \overline{42}_{(i)}| 3\rangle &= J_{(i)}.\\
\end{aligned}
\end{equation}

The expressions for the amplitudes of $\(D^0,D^{+},D^{+}_s\)$ decays into two pseudo-scalar mesons in the framework of broken flavor $SU(3)$ symmetry are rather cumbersome. By making a few assumptions motivated by empirical experience however, one can significantly simplify the task of extracting phenomenology and even making a number of nontrivial predictions out of them. We turn to this task next.

\section{Phenomenology}
\subsection{D Meson Partial Decay Widths}
\label{sec:decaywidth}
The complete decay amplitudes including all matrix elements arising from SU(3) breaking are listed in Appendix \ref{sec:fulldecay}.  In this section we make a simplifying assumption that the matrix elements associated with the three-dimensional representations of  $SU(3)$ in the Hamiltonian are somewhat enhanced compared to higher representations. The assumption can be justified by at least two different lines of reasoning. First, the enhancement of hadronic matrix elements of lower representations 
has been established, albeit not well understood, in neutral kaon systems -- the famous $\Delta=1/2$ rule.
The second, more practical justification is that keeping only the triplet matrix elements from the $SU(3)$ breaking part of the Hamiltonian, one can easily accommodate the known ratio $\Gamma(D^0\to K^+K^-)/\Gamma(D^0\to \pi^+\pi^-)$ and the LHCb measurement of CP asymmetry. Considering more matrix elements will not change this conclusion in any significant way.  As a byproduct, keeping only the triplets from the $SU(3)$ breaking sector will allow us to make non-trivial predictions regarding the direct CP violations in individual $D^0$ decay channels, considered in the next section. In the rest of the paper therefore, we will concentrate on the set of amplitudes consisting of those associated with unbroken $SU(3)$ limit, $(G,F,S,E,T)$, supplemented with the matrix elements corresponding to triplet operators in the Hamiltonian arising from $SU(3)$ breaking, $(F_1,G_1,F_2,G_2,F_3,G_3)$.

Upon a closer inspection of decay amplitudes listed in Appendix \ref{sec:fulldecay}, one can identify a subset in which a relatively small number of linear combinations of different reduced matrix elements is involved. This subset, on which we will concentrate below, includes the following amplitudes,
\begin{align}
\label{eq:decaysimplified}
	\mathcal{A}(D^0\to K^-\pi^+) &= a V^{\phantom{\ast}}_{11}V_{22}^\ast, \nn \displaybreak[0]\\
	\mathcal{A}(D^0\to\bar{K}^0\pi^0) &= \sfrac{-a+5T}{\sqrt{2}}V^{\phantom{\ast}}_{11}V_{22}^\ast,\nn \displaybreak[0]\\
	\mathcal{A}(D^0\to\bar{K}^0\eta) &= \sfrac{-a+5T}{\sqrt{6}}V^{\phantom{\ast}}_{11}V_{22}^\ast,\nn \displaybreak[0]\\
	\mathcal{A}(D^0\to K^+\pi^-) &=  a V^{\phantom{\ast}}_{12}V_{21}^\ast,\nn \displaybreak[0]\\
	\mathcal{A}(D^0\to K^+K^-) &= (a+c)\Sigma + b\Delta,\nn \displaybreak[0]\\
	\mathcal{A}(D^0\to \pi^+\pi^-) &= (-a+c)\Sigma + b\Delta,\nn \displaybreak[0]\\
	\mathcal{A}(D^0\to \pi^0\pi^0) &= \sfrac{-a+ 5T +c}{\sqrt{2}}\Sigma + \sfrac{b-5T}{\sqrt{2}}\Delta,\nn \displaybreak[0]\\
	\mathcal{A}(D^+\to \bar{K}^0\pi^+) &= 5TV^{\phantom{\ast}}_{11}V_{22}^\ast,\nn \displaybreak[0]\\
	\mathcal{A}(D^+\to \pi^+\pi^0) &= \sfrac{5T}{\sqrt{2}}\Sigma - \sfrac{5T}{\sqrt{2}}\Delta,
\end{align}
where we have defined,
\begin{align}
a &= 2T-S+E, \nn\\
b &= \frac{3T+2G+F-E-F_1+\sfrac{2}{3}F_3-2G_1+\sfrac{4}{3}G_3}{2},\nn\\
c &= \frac{F_2+2F_3+2G_2+4G_3}{4}.
\end{align}
The combination $c$ of triplet matrix elements corresponding to the $SU(3)$ breaking part of the Hamiltonian only comes with $\Sigma$, the CKM combination with the large magnitude. As mentioned in Sec.~\ref{sec:intro}, in the SU(3) limit one has to invoke an unacceptably large enhancement of the triplet matrix element coming with $\Delta$ for explaining the ratio $\Gamma(D^0 \to K^+K^-)/\Gamma(D^0 \to \pi^+\pi^-)$. As one can see from \eqref{eq:decaysimplified} however, the $SU(3)$ breaking effects encoded in $c$ can give rise to the difference in the rates for $D^0\to K^+K^-$ and $D^0\to\pi^+\pi^-$ without requiring $b$ to play any r\'ole. 

From the measured values for the partial decay widths of $(D^0,D^+)$ mesons into the final states given above \cite{Nakamura:2010zzi}, one can extract information about the absolute values of the combinations $a$, $c$ and $T$, as well as the magnitudes of the relative phases\footnote{E.g. the decay rate  $\Gamma(D^0\to K^+K^-)\propto |\Sigma|^2|a+c|^2=|\Sigma|^2 (|a|^2+|c|^2+2|a||c|\cos (\phi_c-\phi_a))$, being sensitive only to $|\phi_c-\phi_a|$.}, $|\phi_{ca}| = |\phi_c - \phi_a|$, $|\phi_{Ta}| = |\phi_T - \phi_a|$, where e.g. $a = |a|e^{i\phi_a}$. 

Performing the least $\chi^2$ fit on the corresponding branching ratios, one obtains\footnote{As noted above, we will not need a huge enhancement of $|b|$ with respect to the rest of the amplitudes for accommodating the observations; the partial decay widths for the processes involving both $\Sigma$ and $\Delta$ are therefore dominated by terms proportional to $\Sigma$ in \eqref{eq:decaysimplified}; for extracting information about CP violation however, taking into account the $\Delta$ - contributions is of crucial importance.},
\begin{equation}
\label{eq:chi2fit}
\begin{aligned}
	|a| &= 0.00268\text{ MeV}, \\
	|c| &= 0.00148\text{ MeV}, \\
	|T| &= 0.00029\text{ MeV},
\end{aligned}\quad
\begin{aligned}
	\phi_{ca} &= 0.897,\\
	\phi_{Ta} &= 4.674,
\end{aligned}
\end{equation} 
with the reduced chi-square $\chi_{red}^2\simeq 12$. There is also another solution with the signs of both $\phi_{ca}$ and $\phi_{Ta}$ reversed. We note that $|c|\lsim |a|$ validates our use of $SU(3)$ breaking in organizing the amplitudes, and $|c|/|a|>30\%$ is due to a mild enhancement of the triplet matrix elements.

Notice that knowledge of partial decay widths only allows for the determination of phase differences up to a sign. These signs on the other hand will have effect on CP violation in $D$ decays to be discussed below.
\begin{figure*}
	\subfloat[$\phi_{ca} = 0.897$]{\label{fig:modBpos}\includegraphics[width=0.35\textwidth]{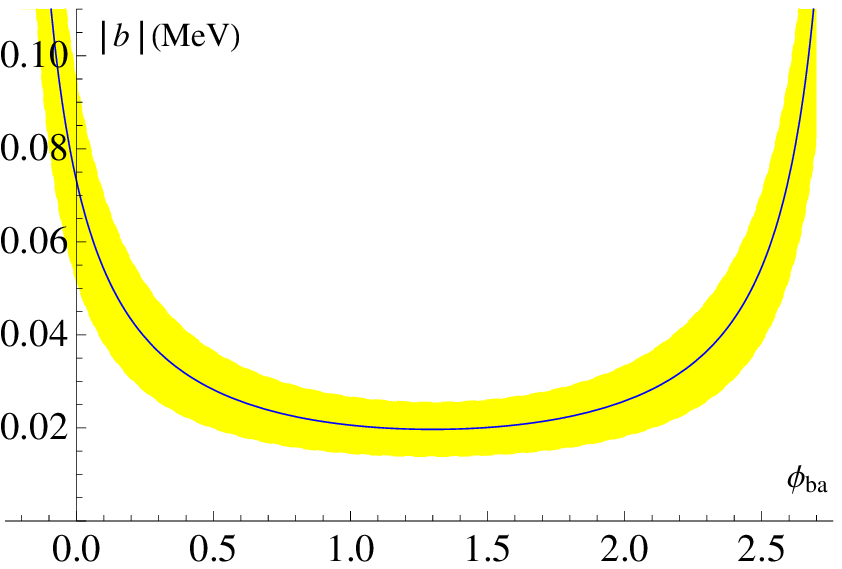}}\qquad
	\subfloat[$\phi_{ca} = -0.897$]{\label{fig:modBneg}\includegraphics[width=0.35\textwidth]{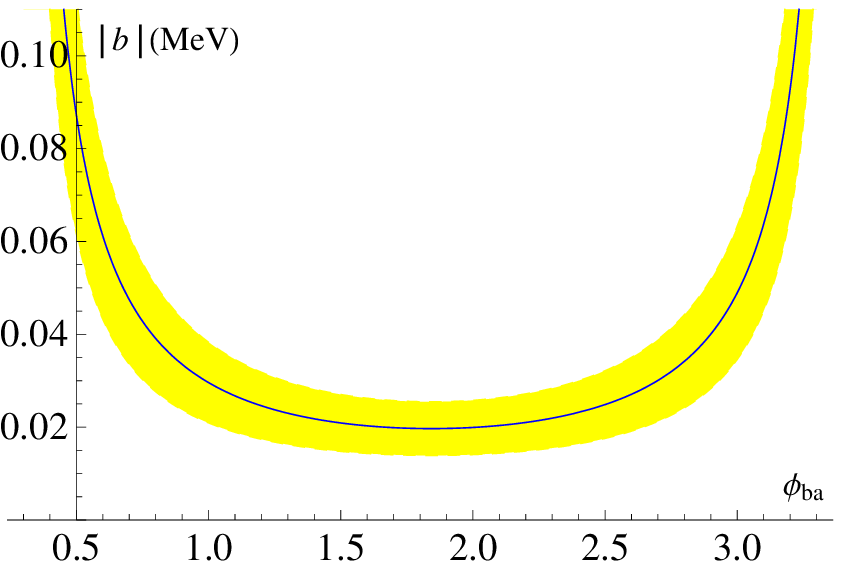}}
	\caption{Values of $|b|$ and $\phi_{ba}$ compatible with $A_{CP}(K^+K^-)-A_{CP}(\pi^+\pi^-)$ reported by the LHCb collaboration. The yellow band represents a 1-$\sigma$ deviation.}
	\label{fig:modB}
\end{figure*}
\subsection{CP Violation in D Decays }
\label{sec:cpv}
Having determined the best-fit values for  
the parameters appearing in \eqref{eq:decaysimplified}, one can proceed to study CP violation in different $D$ decay channels.
Following ref~\cite{Golden:1989qx}, for a decay into two pseudo-scalar mesons $D\to\mathcal{P}\mathcal{P}$ with the amplitude given by
$$\mathcal{A}\( D\to\mathcal{P}\mathcal{P}\)=\tilde{a}\Sigma + \tilde{b}\Delta,$$  
we can write the CP asymmetry as,
\begin{align}
	A_{CP} &= -\frac{2~ \text{Im}(\tilde{a}^\ast\tilde{b})~\text{Im}(\Sigma^\ast\Delta)}{|\tilde{a}|^2|\Sigma|^2 + |\tilde{b}|^2|\Delta|^2+2~\text{Re}(\tilde{a}^\ast\tilde{b})~\text{Re}(\Sigma^\ast\Delta)},\nn\\
	&\approx -2 ~\text{Im}\left(\frac{\tilde{b}}{\tilde{a}}\right)~\text{Im}\left(\frac{\Delta}{\Sigma}\right),
	\label{cpa}
\end{align}
where we have used $|\tilde{a}\Sigma|\gg|\tilde{b}\Delta|$ in the last step. Using the latter equation along with the expressions for the corresponding amplitudes given in (\ref{eq:decaysimplified}), one can determine the one-parameter set of values for $|b|$ and $\phi_{ba}=\phi_b-\phi_a$, compatible with the measurement \eqref{lhcbresult} reported by the LHCb collaboration~\cite{Aaij:2011in} . The results,  including 1$\sigma$ deviation, are given in Fig.~\ref{fig:modB}. Since the observable depends only on the phase difference $\phi_{ca}$, but not on $\phi_{Ta}$, there are two possible cases corresponding to each sign of the former. Recalling the values for the magnitudes of reduced amplitudes given in \eqref{eq:chi2fit}, one can see that a huge enhancement of the combination $b$ (containing unbroken as well as broken $SU(3)$ triplet contributions) with respect to the rest of the amplitudes is not required for accommodating the oberved data. In particular, we will be mostly interested in the range for $b$, corresponding to a factor of $\sim(10-50)$ enhancement with respect to the combination $a$, containing the unbroken $SU(3)$ - limit matrix elements $(T,S,E)$. As mentioned above, such an enhancement is motivated by the $\Delta=1/2$ rule in Kaon decays.

\begin{figure*}
	\subfloat[$\phi_{ca} > 0$, $\phi_{Ta} > 0$]{\label{fig:Acp1}\includegraphics[width=0.35\textwidth]{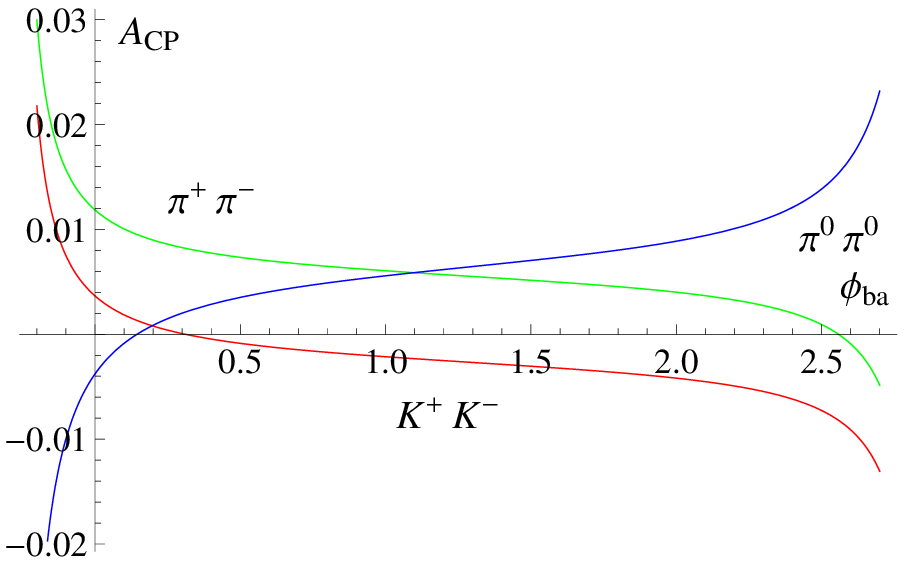}}\qquad
	\subfloat[$\phi_{ca} < 0$, $\phi_{Ta} < 0$]{\label{fig:Acp4}\includegraphics[width=0.35\textwidth]{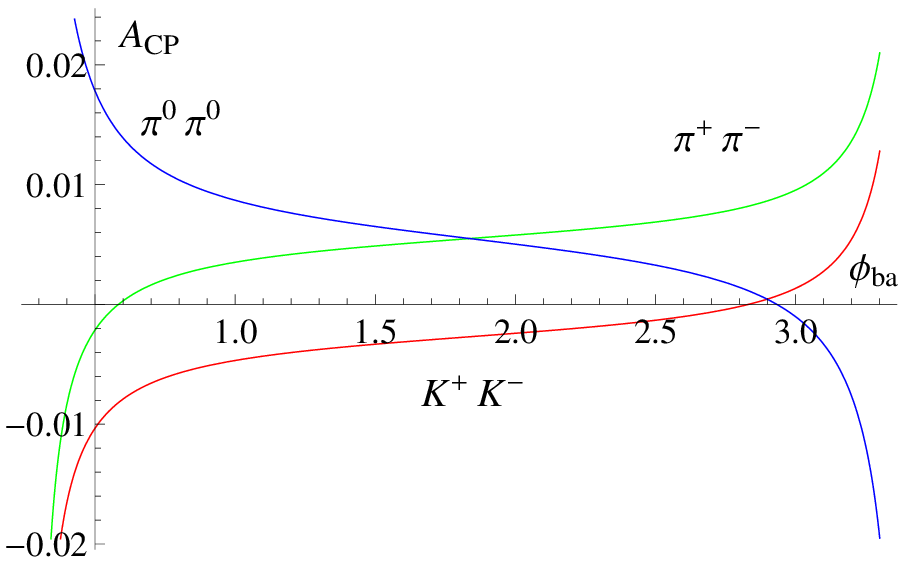}}
	\caption{CP asymmetry in $K^+K^-$ (red), $\pi^+\pi^-$ (green) and $\pi^0\pi^0$ (blue) channels. The central values of asymmetries shown here correspond to $A_{CP}(K^+K^-)-A_{CP}(\pi^+\pi^-)$ reported by the LHCb collaboration.}
	\label{fig:Acp}
\end{figure*}

We are now ready to discuss direct CP violation in individual D meson decay modes. Since the expression for CP asymmetry given in \eqref{eq:chi2fit} is sensitive to the common sign of $\phi_{ca}$ and $\phi_{Ta}$, one should consider two cases, corresponding to each of the possible signs. The dependence of individual CP asymmetries $A_{CP}(D^0\to K^+K^-)$, $A_{CP}(D^0\to\pi^+\pi^-)$ and $A_{CP}(D^0\to\pi^0\pi^0)$ on the relative phase $\phi_{ba}$ 
is shown for each of the two possibilities in Fig. \ref{fig:Acp}. We only plot central values for asymmetries in order not to overload the figures, while keeping the 1$\sigma$ deviation does not change results in any qualitative way.

There are a number of observations we can make from these plots. First of all, one can see that depending on the amount of enhancement of the triplet operators, $|A_{CP}(D^0\to\pi^0\pi^0)|$ can be as large as a few percent. We also expect $|A_{CP}(D^0\to K^+K^-)-A_{CP}(D^0\to\pi^0\pi^0)|$ to exceed $|A_{CP}(D^0\to K^+K^-)-A_{CP}(D^0\to\pi^+\pi^-)|$ for a large part of parameter space.

Moreover, from a precise measurement of individual asymmetries in decays into $K^+K^-$ and $\pi^+\pi^-$, one can  predict the value of $A_{CP}(D^0\to\pi^0\pi^0)$. CDF collaboration has recently reported $A_{CP}(K^+K^-)=-0.24 \pm 0.24 \%$ and $A_{CP}(\pi^+\pi^-)= 0.22 \pm 0.26 \%$~\cite{Aaltonen:2011se}, which are consistent with most of the parameter space, considered in Fig. \ref{fig:Acp}. Depending on the particular values of the phases $\phi_{ca}$ and $\phi_{Ta}$, this would imply $A_{CP}(\pi^0\pi^0)\sim (0.5 - 1) \%$, as seen from the two plots in Fig. \ref{fig:Acp}. The $\pi^0\pi^0$ final state is difficult to observe experimentally; however, our framework makes it possible to make predictions about asymmetries in other pseudo-scalar final states (including the effects of $\eta-\eta'$ mixing), which can be more easily tested. The complete analysis is in progress and will be addressed in a separate publication.

Note that the asymmetry $A_{CP}(D^+\to\pi^+\pi^0)$ vanishes. This persists even if all $SU(3)$ breaking matrix elements are included as can be seen from Appendix~\ref{sec:fulldecay}. 


\begin{acknowledgements}
We thank Ben Grinstein for pointing out the problem, many enlightening discussions and comments on the manuscript. We also thank Diego Guadagnoli and the anonymous referees for valuable comments on the manuscript. This work was supported by the  US Department of Energy under contract DOE-FG03-97ER40546.
\end{acknowledgements}

\appendix
\section{Full SU(3) Breaking Decay Amplitudes}
\label{sec:fulldecay}
In this appendix we collect the complete expressions for the amplitudes of $(D^0,D^+,D^+_s)$ decays into two pseudo-scalar final states to first order in $SU(3)$ breaking. The notation for matrix elements of different $SU(3)$ representations of the weak Hamiltonian is defined in \eqref{matrixelements}. 
The notation for $SU(3)$ - limit matrix elements agrees with that of Quigg \cite{Quigg:1979ic}\footnote{Note that our $SU(3)$ symmetry - limit results given below exactly agree with those of \cite{Quigg:1979ic}, up to a convention of multiplying all same-particle final state amplitudes by a factor of $\sqrt{2}$, which we use here.}.
\vskip 0.3cm
\textbf{Amplitudes for $D^0$ decays}
\begin{align}
	\scriptstyle K^-\pi^+&:\textstyle{(}\scriptstyle{2T + E - S}+\sfrac{3 E_2+2E_3^a-4E_3^b}{4}+\sfrac{2H_2-H_3}{20}\nn \\ \displaybreak[0]
	&\scriptstyle{-\sfrac{J_3}{3}+\sfrac{2S_2+3 S_3}{2}+\sfrac{3 T_2+2T_3^a-4T_3^b}{2}\textstyle{)}\;V^{\phantom{\ast}}_{11}V_{22}^\ast}\nn \\ 
	\scriptstyle\bar{K}^0\pi^0&:\textstyle{(}\scriptstyle{3T - E + S}-\sfrac{3 E_2+2E_3^a-4E_3^b}{4}-\sfrac{6 H_2-3H_3}{10}\nn \\ \displaybreak[0]
	&\scriptstyle{ -\sfrac{2S_2+3 S_3}{2}+\sfrac{9 T_2+6 T_3^a-12T_3^b}{4}\textstyle{)}}\sfrac{V^{\phantom{\ast}}_{11}V_{22}^\ast}{\sqrt{2}}\nn \\
	\scriptstyle \bar{K}^0\eta &: \textstyle{(}\scriptstyle 3T - E + S-\sfrac{3 E_2+2E_3^a-4E_3^b}{4}+\sfrac{12 H_2-6 H_3}{5}\nn \\ \displaybreak[0]
	\scriptstyle&\scriptstyle{-2 J_3}-\sfrac{2S_2+3 S_3}{2}+\sfrac{9 T_2+6 T_3^a-12T_3^b}{4}\textstyle{)}\sfrac{V^{\phantom{\ast}}_{11}V_{22}^\ast}{\sqrt{6}}\nn \\
	\scriptstyle K^+\pi^- &: \textstyle{(}\scriptstyle{2T + E - S-E_3^a+2 E_3^b}+\sfrac{6 H_2+5H_3}{20}\nn \\ \displaybreak[0]
	&\scriptstyle{-\sfrac{J_3}{3}-2 S_2-2 T_3^a+4 T_3^b}\textstyle{)}\; V^{\phantom{\ast}}_{12}V_{21}^\ast\nn \\
	\scriptstyle K^0\pi^0&: \textstyle{(} \scriptstyle 3T - E + S+E_3^a-2 E_3^b-\sfrac{3 H_2}{10}+\sfrac{3 H_3}{4}\nn \\  \displaybreak[0]
	& \scriptstyle +2 S_2-3 T_3^a+6 T_3^b\textstyle{)}\sfrac{V^{\phantom{\ast}}_{12}V_{21}^\ast}{\sqrt{2}}\nn \\
	\scriptstyle K^0\eta &: \textstyle{(}\scriptstyle 3T - E + S+E_3^a-2 E_3^b+\sfrac{27 H_2}{10}-\sfrac{3 H_3}{4}\nn \\ \displaybreak[0]
	&\scriptstyle -2 J_3+2 S_2-3 T_3^a+6 T_3^b\textstyle{)}\sfrac{V^{\phantom{\ast}}_{12}V_{21}^\ast}{\sqrt{6}}\nn \\
	\scriptstyle K^+K^- &: \textstyle{(}\scriptstyle2T+E-S +\sfrac{3 E_2}{16}-\sfrac{5 E_3^a}{8}-\sfrac{E_3^b}{4}+\sfrac{F_2+2F_3}{4}\nn\\
	&\quad\scriptstyle+\sfrac{G_2+2G_3}{2}+\sfrac{H_2}{5}+\sfrac{H_3}{10}-\sfrac{13 J_3}{30}-\sfrac{S_2}{2}+\sfrac{3 S_3}{4}\nn \\
	&\quad\scriptstyle+\sfrac{21 T_2}{16}+\sfrac{5 T_3^a}{8}+\sfrac{13 T_3^b}{4}\textstyle{)}\Sigma \nn\displaybreak[0] \\
	&\scriptstyle+ \sfrac{1}{2}\textstyle{(}\scriptstyle3T+2G+F-E+\sfrac{9 E_1}{4}+E_3^a+E_3^b-F_1+\sfrac{2 F_3}{3}\nn \\ 
	&\quad\scriptstyle-2 G_1+\sfrac{4 G_3}{3}-\sfrac{2 H_3}{5}-\sfrac{19 J_3}{15}+\sfrac{9 S_1}{2}\nn\\
	&\quad\scriptstyle+\sfrac{63 T_1}{4}-3 T_3^a-3 T_3^b\textstyle{)}\Delta\nn \displaybreak[0]\\
	\scriptstyle \pi^+\pi^-&: \scriptstyle-\textstyle{(}\scriptstyle2T+E-S +\sfrac{9 E_2}{16}+\sfrac{E_3^a}{8}+\sfrac{5 E_3^b}{4}-\sfrac{F_2}{4}-\sfrac{F_3}{2}\nn \\
	&\qquad\scriptstyle-\sfrac{G_2}{2}-G_3+\sfrac{H_2}{5}+\sfrac{H_3}{10}-\sfrac{7 J_3}{30}-\sfrac{S_2}{2}+\sfrac{3 S_3}{4}\nn\\
	&\qquad\scriptstyle+\sfrac{3 T_2}{16}-\sfrac{13 T_3^a}{8}-\sfrac{5 T_3^b}{4}\textstyle{)}\Sigma \nn  \displaybreak[0] \\
	&\scriptstyle+ \sfrac{1}{2}\textstyle{(}\scriptstyle3T+2G+F-E-\sfrac{27 E_1}{4}+E_3^a+E_3^b-F_1+\sfrac{2 F_3}{3}\nn \\
	&\qquad\scriptstyle-2 G_1+\sfrac{4 G_3}{3}+\sfrac{2 H_3}{5}+\sfrac{J_3}{15}-\sfrac{9 S_1}{2}\nn\\
	&\qquad\scriptstyle-\sfrac{9 T_1}{4}-3 T_3^a-3 T_3^b\textstyle{)}\Delta\nn  \displaybreak[0]\\
	\scriptstyle\pi^0\pi^0 &: \sfrac{1}{\sqrt{2}}\textstyle{(}\scriptstyle3T-E+S +\sfrac{9 E_2}{16}+\sfrac{E_3^a}{8}+\sfrac{5 E_3^b}{4}-\sfrac{F_2}{4}-\sfrac{F_3}{2}\nn\\
	&\qquad\scriptstyle-\sfrac{G_2}{2}-G_3+\sfrac{H_2}{5}-\sfrac{2 H_3}{5}-\sfrac{J_3}{15}-\sfrac{S_2}{2}+\sfrac{3 S_3}{4}\nn \\
	&\qquad\scriptstyle+\sfrac{3 T_2}{16}+\sfrac{27 T_3^a}{8}+\sfrac{15 T_3^b}{4}\textstyle{)}\Sigma\nn\displaybreak[0] \\
	&\scriptstyle+ \sfrac{1}{2\sqrt{2}}\textstyle{(}\scriptstyle-7T+2G+F-E-\sfrac{27 E_1}{4}+E_3^a+E_3^b\nn\\
	&\qquad\scriptstyle-F_1+\sfrac{2 F_3}{3}-2 G_1+\sfrac{4 G_3}{3}-\sfrac{3 H_3}{5}+\sfrac{2 J_3}{5}\nn \\
	&\qquad\scriptstyle-\sfrac{9 S_1}{2}-\sfrac{9 T_1}{4}+7 T_3a+7 T_3b\textstyle{)}\Delta\nn \displaybreak[0]\\ 
	\scriptstyle K^0\bar{K}^0 &: \textstyle{(}\scriptstyle\sfrac{3 E_2}{8}+\sfrac{3 E_3^a}{4}+\sfrac{3 E_3^b}{2}-\sfrac{F_2}{2}-F_3+\sfrac{G_2}{2}+G_3\nn \\
	&\qquad\scriptstyle +H_2-\sfrac{H_3}{2}-\sfrac{19 J_3}{30}-\sfrac{3 T_2}{16}-\sfrac{3 T_3^a}{8}-\sfrac{3 T_3^b}{4}\textstyle{)}\Sigma \nn\displaybreak[0] \\
	&\scriptstyle+ \textstyle{(}\scriptstyle-\sfrac{T}{2}+G-F+E+\sfrac{9 E_1}{4}-E_3^a-E_3^b+F_1-\sfrac{2 F_3}{3}\nn\\ 
	&\qquad\scriptstyle-G_1+\sfrac{2 G_3}{3}+\sfrac{J_3}{10}-\sfrac{9 T_1}{8}+\sfrac{T_3^a}{2}+\sfrac{T_3^b}{2}\textstyle{)}\Delta\nn\displaybreak[0] \\
	\scriptstyle\eta\eta &: \scriptstyle \sfrac{1}{\sqrt{2}}\textstyle{(}\scriptstyle-3T+E-S +\sfrac{9 E_2}{16}+\sfrac{E_3^a}{8}+\sfrac{5 E_3^b}{4}-\sfrac{F_2}{4}-\sfrac{F_3}{2}\nn\\
	&\qquad\scriptstyle+\sfrac{G_2}{2}+G_3-\sfrac{9 H_2}{5}+\sfrac{3 H_3}{5}+\sfrac{8 J_3}{5}-\sfrac{S_2}{2}+\sfrac{3 S_3}{4}\nn \\
	&\qquad\scriptstyle-\sfrac{27 T_2}{16}-\sfrac{3 T_3^a}{8}-\sfrac{15 T_3^b}{4}\textstyle{)}\Sigma\nn \displaybreak[0]\\
	&\scriptstyle + \sfrac{1}{2\sqrt{2}}\textstyle{(}\scriptstyle-3T+2G-F+E+\sfrac{27 E_1}{4}-E_3^a-E_3^b\nn\\
	&\qquad\scriptstyle+F_1-\sfrac{2 F_3}{3}-2 G_1+\sfrac{4 G_3}{3}+\sfrac{3 H_3}{5}+\sfrac{8 J_3}{5}\nn \\ 
	&\qquad\scriptstyle+\sfrac{9 S_1}{2}-\sfrac{81 T_1}{4}+3 T_3^a+3 T_3^b\textstyle{)}\Delta\nn \\ \displaybreak[0]
	\scriptstyle\eta\pi^0&:\scriptstyle -\sfrac{1}{\sqrt{3}}\textstyle{(}\scriptstyle3T-E+S+ \sfrac{3 E_2}{16}+\sfrac{11 E_3^a}{8}+\sfrac{7 E_3^b}{4}-\sfrac{3 F_2}{4}-\sfrac{3 F_3}{2}\nn \\
	&\quad\scriptstyle-\sfrac{6 H_2}{5}+\sfrac{9 H_3}{10}+\sfrac{3 J_3}{10}+\sfrac{S_2}{2}-\sfrac{3 S_3}{4}+\sfrac{9 T_2}{4}+\sfrac{3 T_3^a}{2}+6 T_3^b\textstyle{)}\Sigma\nn \\
	&\scriptstyle +\sfrac{\sqrt{3}}{2} \textstyle{(}\scriptstyle-2T+F-E-\sfrac{3 E_1}{4}+E_3^a+E_3^b-F_1+\sfrac{2 F_3}{3}\nn \\
	&\qquad\scriptstyle+\sfrac{H_3}{5}+\sfrac{11 J_3}{15}+\sfrac{3 S_1}{2}-9 T_1+2 T_3a+2 T_3b\textstyle{)}\Delta\nn
\end{align}
\textbf{Amplitudes for $D^+$ decays}
\begin{align}
	\scriptstyle \bar{K}^0\pi^+ &: \textstyle{(}\scriptstyle5T-\sfrac{H_2}{2}+\sfrac{H_3}{4}-\sfrac{J_3}{3}+\sfrac{15 T_2}{4}+\sfrac{5 T_3^a}{2}-5 T_3^b\textstyle{)}\;V^{\phantom{\ast}}_{11}V_{22}^\ast\nn \\ \displaybreak[0]
	\scriptstyle K^0\pi^+ &: \textstyle{(}\scriptstyle2T+E+S-E_3^a+2 E_3^b-\sfrac{3 H_2}{10}+\sfrac{H_3}{4}-\sfrac{J_3}{3}+2 S_2\nn\\
	&\qquad\scriptstyle -2 T_3^a+4 T_3^b\textstyle{)}\;V^{\phantom{\ast}}_{12}V_{21}^\ast\nn  \displaybreak[0] \\ 
	\scriptstyle K^+\pi^0 &: \textstyle{(}\scriptstyle3T - E - S+E_3^a-2 E_3^b+\sfrac{3 H_2}{10}+\sfrac{3 H_3}{4}-2 S_2\nn\\
	&\qquad\scriptstyle-3 T_3^a+6 T_3^b\textstyle{)}\sfrac{V^{\phantom{\ast}}_{12}V_{21}^\ast}{\sqrt{2}}\nn \displaybreak[0] \\ 
	\scriptstyle K^+\eta &: \textstyle{(}\scriptstyle3T - E - S+E_3^a-2 E_3^b-\sfrac{27 H_2}{10}-\sfrac{3 H_3}{4}-2 J_3\nn\\
	&\qquad\scriptstyle-2 S_2-3 T_3^a+6 T_3^b\textstyle{)}\sfrac{V^{\phantom{\ast}}_{12}V_{21}^\ast}{\sqrt{6}}\nn \displaybreak[0]\\ 
	\scriptstyle \pi^0\pi^+&: \sfrac{1}{\sqrt{2}}\textstyle{(}\scriptstyle5T+\sfrac{H_3}{2}-\sfrac{J_3}{6}-5 T_3^a-5 T_3^b\textstyle{)}\Sigma\nn\\ \displaybreak[0]
	&\scriptstyle +\sfrac{1}{\sqrt{2}}\textstyle{(}\scriptstyle-5T-\sfrac{H_3}{2}+\sfrac{J_3}{6}+5 T_3^a+5 T_3^b\textstyle{)}\Delta\nn\\
	\scriptstyle \pi^+\eta &: -\sfrac{1}{\sqrt{6}}\textstyle{(}\scriptstyle9T+2E+2S +\sfrac{3 E_2}{8}-\sfrac{5 E_3^a}{4}-\sfrac{E_3^b}{2}-\sfrac{3 F_2}{2}-3 F_3\nn\\
	&\qquad\scriptstyle-\sfrac{12 H_2}{5}+\sfrac{3 H_3}{10}-\sfrac{19 J_3}{10}+S_2-\sfrac{3 S_3}{2}+\sfrac{9 T_2}{2}+9 T_3^b\textstyle{)}\Sigma\nn\\
	&\scriptstyle+ \sfrac{1}{\sqrt{6}}\textstyle{(}\scriptstyle-3T+3F+E-\sfrac{9 E_1}{4}-E_3^a-E_3^b-3 F_1+2 F_3-\sfrac{9 H_3}{10}\nn\\ \displaybreak[0]
	&\qquad\scriptstyle-\sfrac{3 J_3}{10}+\sfrac{9 S_1}{2}-27 T_1+3 T_3^a+3 T_3^b\textstyle{)}\Delta\nn\\ 
	\scriptstyle K^+\bar{K}^0 &: \textstyle{(}\scriptstyle3T-E-S -\sfrac{3 E_2}{16}+\sfrac{5 E_3^a}{8}+\sfrac{E_3^b}{4}+\sfrac{3 F_2}{4}+\sfrac{3 F_3}{2}\nn\\
	&\qquad\scriptstyle-\sfrac{4 H_2}{5}+\sfrac{H_3}{10}-\sfrac{19 J_3}{30}-\sfrac{S_2}{2}+\sfrac{3 S_3}{4}+\sfrac{3 T_2}{2}+3 T_3^b\textstyle{)}\Sigma\nn\displaybreak[0]\\
	&\scriptstyle+ \sfrac{1}{2}\textstyle{(}\scriptstyle2T+3F+E-\sfrac{9 E_1}{4}-E_3^a-E_3^b-3 F_1+2 F_3+\sfrac{3 H_3}{5}\nn\\
	&\qquad\scriptstyle+\sfrac{J_3}{5}+\sfrac{9 S_1}{2}+18 T_1-2 T_3^a-2 T_3^b\textstyle{)}\Delta\nn
\end{align}
\textbf{Amplitudes for $D^+_s$ decays}
\begin{align}
	\scriptstyle \bar{K}^0K^+ &: \textstyle{(}\scriptstyle2T+E+S+\sfrac{3 E_2}{4}+\sfrac{E_3^a}{2}-E_3^b+\sfrac{2 H_2}{5}-\sfrac{H_3}{5}+\sfrac{2 J_3}{3}\nn\\ 
	&\qquad\scriptstyle-S_2-\sfrac{3 S_3}{2}+\sfrac{3 T_2}{2}+T_3^a-2 T_3^b\textstyle{)}\;V^{\phantom{\ast}}_{11}V_{22}^\ast\nn \displaybreak[0] \\ 
	\scriptstyle \eta\pi^+ &: \sqrt{\sfrac{2}{3}}\textstyle{(}\scriptstyle-3T+E+S+\sfrac{3 E_2}{4}+\sfrac{E_3^a}{2}-E_3^b-\sfrac{3 H_2}{5}+\sfrac{3 H_3}{10}\nn\\ \displaybreak[0]
	&\qquad\scriptstyle-J_3-S_2-\sfrac{3 S_3}{2}-\sfrac{9 T_2}{4}-\sfrac{3 T_3^a}{2}+3 T_3^b\textstyle{)}\;V^{\phantom{\ast}}_{11}V_{22}^\ast\nn \\ 
	\scriptstyle K^0K^+ &: \textstyle{(}\scriptstyle5T-\sfrac{H_3}{2}+\sfrac{2 J_3}{3}-5 T_3^a+10 T_3^b\textstyle{)}\;V^{\phantom{\ast}}_{12}V^\ast_{21}\nn \\ \displaybreak[0]
	\scriptstyle K^0\pi^+ &: \textstyle{(}\scriptstyle-3T+E+S +\sfrac{9 E_2}{16}+\sfrac{E_3^a}{8}+\sfrac{5 E_3^b}{4}+\sfrac{3 F_2}{4}+\sfrac{3 F_3}{2}\nn\\
	&\qquad\scriptstyle-\sfrac{H_2}{5}+\sfrac{2 H_3}{5}-\sfrac{19 J_3}{30}+\sfrac{S_2}{2}-\sfrac{3 S_3}{4}-\sfrac{3 T_2}{4}+\sfrac{3 T_3^a}{2}\textstyle{)}\Sigma\nn\\
	&\scriptstyle + \sfrac{1}{2}\textstyle{(}\scriptstyle2T+3F+E+\sfrac{27 E_1}{4}-E_3^a-E_3^b-3 F_1+2 F_3-\sfrac{3 H_3}{5}\nn\\ 
	&\qquad\scriptstyle +\sfrac{J_3}{5}-\sfrac{9 S_1}{2}-9 T_1-2 T_3^a-2 T_3^b\textstyle{)}\Delta\nn\displaybreak[0] \\ 
	\scriptstyle K^+\pi^0 &: \sfrac{1}{\sqrt{2}}\textstyle{(}\scriptstyle2T+E+S+ \sfrac{9 E_2}{16}+\sfrac{E_3^a}{8}+\sfrac{5 E_3^b}{4}+\sfrac{3 F_2}{4}+\sfrac{3 F_3}{2}-\sfrac{H_2}{5}\nn\\
	&\qquad\scriptstyle-\sfrac{3 H_3}{5}+\sfrac{J_3}{5}+\sfrac{S_2}{2}-\sfrac{3 S_3}{4}-\sfrac{3 T_2}{4}-\sfrac{7 T_3^a}{2}-5 T_3^b\textstyle{)}\Sigma\nn\displaybreak[0]\\
	&\scriptstyle + \sfrac{1}{\sqrt{2}}\textstyle{(}\scriptstyle-4T+\sfrac{3F}{2}+\sfrac{E}{2}+\sfrac{27 E_1}{8}-\sfrac{E_3^a}{2}-\sfrac{E_3^b}{2}-\sfrac{3 F_1}{2}+F_3\nn\\  
	&\qquad\scriptstyle +\sfrac{7 H_3}{10}-\sfrac{11 J_3}{15}-\sfrac{9 S_1}{4}-\sfrac{9 T_1}{2}+4 T_3^a+4 T_3^b\textstyle{)}\Delta \nn\displaybreak[0] \\
	\scriptstyle K^+\eta &: -\sfrac{1}{\sqrt{6}}\textstyle{(}\scriptstyle12T+E+S +\sfrac{9 E_2}{16}+\sfrac{E_3^a}{8}+\sfrac{5 E_3^b}{4}+\sfrac{3 F_2}{4}+\sfrac{3 F_3}{2}\nn\\
	&\qquad\scriptstyle+\sfrac{9 H_2}{5}-\sfrac{3 H_3}{5}+\sfrac{16 J_3}{5}+\sfrac{S_2}{2}-\sfrac{3 S_3}{4}\nn\\
	&\qquad\scriptstyle+\sfrac{27 T_2}{4}+\sfrac{3 T_3^a}{2}+15 T_3^b\textstyle{)}\Sigma \nn\displaybreak[0]\\
	&\scriptstyle- \sfrac{1}{\sqrt{6}}\textstyle{(}\scriptstyle6T+\sfrac{3F}{2}+\sfrac{E}{2}+\sfrac{27 E_1}{8}-\sfrac{E_3^a}{2}-\sfrac{E_3^b}{2}-\sfrac{3 F_1}{2}+F_3\nn\\
	&\qquad\scriptstyle-\sfrac{3 H_3}{10}+\sfrac{8 J_3}{5}-\sfrac{9 S_1}{4}+\sfrac{81 T_1}{2}-6 T_3^a-6 T_3^b\textstyle{)}\Delta\nn
\end{align}


\bibliography{ddecay}{}
\bibliographystyle{apsrev4-1}

\end{document}